\documentstyle[12pt]{article}
\def\Box{\hbox{$\sqcup$\kern-0.66em\lower0.03ex\hbox{$\sqcap$}}}
\begin{document}
\begin{titlepage}
\begin{flushright}
IFUP--TH 68/99 \\
LPT-ENS-99/59\\
\bigskip
\end{flushright}
\vskip 1truecm
\begin{center}
\Large\bf
ADM approach to 2+1 dimensional gravity
\footnote{Contribution to Third Meeting on Constrained Dynamics
and Quantum Gravity, Villasimius,  
September 13-17, 1999 }
\end{center}
\vskip 1truecm
\begin{center}
{Pietro Menotti} \footnote{This work is  supported in part
  by M.U.R.S.T.}
\\ 
{\small\it Dipartimento di Fisica dell'Universit{\`a}, Pisa 56100, 
Italy and}\\
{\small\it INFN, Sezione di Pisa}\\
\end{center}
\vskip .8truecm
\begin{center}
{Domenico Seminara\footnote{CEE Post-doctoral Fellow under contract
FMRX CT96-0045 }} \\  
{\small\it Laboratoire de Physique Th{\'e}orique, 
{\'E}cole Normale Sup{\'e}rieure\footnote{Unit{\'e} Mixte associee au
Centre de la Recherche Scientifique et  \'a l' \'Ecole Normale
Sup\'erieure.},\\
F-75231, Paris CEDEX 05, France}\\
\end{center}
\begin{center}
December 1999
\end{center}
\end{titlepage}

\begin{abstract} 
The canonical ADM equations are solved in terms of
the conformal factor in the instantaneous York gauge. A simple
derivation is given for the solution of the two body problem.  A
geometrical characterization is given for the apparent singularities
occurring in the ${\cal N}$-body problem and it is shown how the
Garnier hamiltonian system arises in the ADM treatment by considering
the time development of the conformal factor at the locations where
the extrinsic curvature tensor vanishes. The equations of motion for
the position of the particles and of the apparent singularities and
also the time dependence of the linear residues at such singularities
are given by the transformation induced by an energy momentum tensor
of a conformal Liouville theory. Such an equation encodes completely
the dynamics of the system.  
\end{abstract}


\section{Introduction}

The problem of providing solutions to $2+1$ dimensional gravity \cite
{DJH} has been approached along different lines. In absence of
particles the dynamics becomes non trivial only for closed universes
of genus higher or equal to one. Moncrief \cite{moncrief} and Hosoya
and Nakao \cite{hosoya} gave the hamiltonian treatment of this problem
providing the complete reduction of the hamiltonian to the physical
parameters i.e. the moduli of the time slices. This procedure was
obtained in the York $K=t$ gauge; in the case of
the torus the problem can be dealt with explicitly while for higher
genus, even if the problem is well defined we do
not posses an explicit form for the reduced hamiltonian.

A different approach was put forward by 't Hooft \cite{hooft} by
describing the evolving Cauchy surfaces in terms of polygonal tiles
which join along segments where the extrinsic curvature is
singular. The dynamics of the system is codified in transition rules
which intervene when e.g. the length of a side of a polygon goes to
zero or when a particle collides with a  side of a polygon. Such an
approach is applicable both in presence and in absence of particle and
it is suitable for an algorithmic or numeric approach. A similar
but different approach was given by  Waelbroek \cite{wael}.

In presence of particles progress was made in the papers by Bellini,
Ciafaloni and Valtancoli \cite{BCV} and by Welling \cite{welling} in
the first order formalism by going over to the instantaneous ($K=0$)
gauge; in
these papers it was shown that the problem is equivalent to a Riemann-
Hilbert problem. It can be solved explicitly 
for the case of two particles, in terms of hypergeometric
functions. For three or more particles one encounters a feature known
in the mathematical literature as apparent singularities. 

In this paper we shall consider $2+1$ dimensional gravity in presence
of massive particles by exploiting the hamiltonian formulation. Thus
our setting is a second order 
formalism and the basic equations are those derived systematically
from the variation of the ADM action. The reason for such a
development is to give a treatment which resides completely within the
canonical framework. All results obtained in
\cite{BCV} \cite{welling} come out in simple fashion from such an
approach; moreover we shall prove that the Garnier hamiltonian system
for the apparent singularities is a direct outcome of the canonical
ADM formalism.  
Again the
exploitation of the instantaneous York gauge plays a major role; the
technical advantage of such a gauge is to reduce an equation of
sinh-Gordon to one of Liouville type to which powerful methods of
complex analysis apply.  Unfortunately the applicability of such a
gauge is restricted to open universes.

We shall see that the described approach provides an elementary
way to solve the two body problem: the exact solution of the motion
can be derived even before the explicit computation of the metric;
combined with the qualitative knowledge of the asymptotic metric it
gives a complete description of the scattering. The exact metric is
obtained by solving a Liouville equation, leading of course to the
same results obtained in \cite{BCV} and \cite{welling}.

One of the aims of presenting a canonical solution of the problem is
to provide a framework for the quantization of the theory even if here
we shall confine ourselves to the classical problem.

Quantization schemes for $2+1$ dimensional gravity in absence of
particles have been proposed in \cite{nelsonregge,carlip,hosoya2}
and in presence of particles in \cite{hooft2,waelbroek2,matschull}. 

The key quantity of our approach will be the conformal factor which
describes the metric of the time slices. The lapse function $N$ is
given by the derivative of such conformal factor with respect to the
total energy and the shift functions are given in terms of derivatives
of the lapse function.

As mentioned above, with more than two particles there are 
apparent singularities and one has to provide evolution equations for
their positions and residues. We show that all this information in
contained in the ADM equation for the time derivative of the
conformal factor. In fact by exploiting Schwarz's relation between the
coefficient of the fuchsian differential equation and the reduced
conformal factor one obtains the Garnier equation by equating the
residues at the polar singularities at the position of the apparent
singularities.

In the present paper we shall outline the main features of the
approach; technical details are found in ref.\cite{pmdsann}.

\section{The action}

The action for the gravitational field including the boundary terms is
given by \cite{hawkinghunter} 
$$ \frac{ 16 \pi
G_N}{c^3}S_{Grav}=\int_{{\cal M}}dt~ d^2 x 
\sqrt{\rm {\bf g}}~ {\cal R} + 
$$ 
\begin{equation}\label{action}
2\int_{\Sigma_0}^{\Sigma_1}d^2x \sqrt{\bf g}\,K + 
2\int_B d^2x \sqrt{-\gamma}\,\Theta +
\end{equation} 
$$
2\int_{B_0}^{B_1}d x
\sqrt{\sigma}~ \eta 
$$ 
where ${\bf g}_{\mu \nu}$, ${\cal R}$ are the
$2+1$ dimensional metric and curvature, $g_{ij}$ the $2$-dimensional
metric of the constant time slices $\Sigma_t$; $\Sigma_0$ and
$\Sigma_1$ the initial and final time slices; $K_{ij}$ the second
fundamental form of the time slices $\Sigma_t$ and $\Theta_{ij}$ the
second fundamental form of the lateral boundary $B$ whose volume form
is $\sqrt{-\gamma}$; $n^\mu $ is the future pointing unit normal to
the time slices and $u^\mu $ the outward pointing unit normal to $B$;
$B_t= \Sigma_t\cap B$, $\sqrt{\sigma}$ is the volume form induced on
$B_t$ and $\sinh\eta=n_\mu u^\mu $.
The boundary terms are going to play a fundamental role in the
computation of the reduced action; this paper however will be based
completely on the equations of motion which do not depend on the
explicit form of the boundary terms.

In presence of particles we must add to (\ref{action}) the term 
\begin{equation}
-\int dt \sum_n m_n\sqrt{-{\bf g}_{\mu
\nu}(q_n)\dot q_n^\mu \dot q_n^\nu} 
\end{equation} 
with $m_n = \frac{16 \pi G_N {\rm m}_n}{c^2}$,  $0<m_n<4\pi$.
In the ADM notation where the metric is written as \cite{adm}
\begin{equation} 
ds^2 = - N^2 dt^2 +g_{ij}(dx^i+N^idt)(dx^j+N^jdt)
\end{equation} 
the volume term apart for a divergence, can be rewritten
in terms of 
the canonical variables as \cite{adm,hawkinghunter,wald}  
\begin{equation} 
\int_{{\cal M}}  dt~d^2x 
\left[ \pi^{ij}\dot g_{ij} - N^i H_i - NH\right]
\end{equation} 
where we have introduced the momenta $\pi^{ij}$ canonically conjugate to
$g_{ij}$. Similarly the matter action takes the form
\begin{equation} 
\int\!d t \sum_n\Big(P_{ni}\, \dot
q_n^i+N^i(q_n) P_{ni} - 
\end{equation}
$$
N(q_n) \sqrt{P_{ni} P_{nj} g^{ij}(q_n)+
m_n^2}\Big). 
$$ 

\section{The constraints}
In $(2+1)-$dimensions, 
the great advantage of a York-type gauge, $\partial_i K=0$, is to ensure,
when combined with the conformal gauge $g_{ij}=\exp(2\sigma)\delta_{ij}$,
the decoupling of the traceless part $\pi^{Tij}$ of the momentum from
the spatial  
metric in the diffeomorphism constraints\footnote{This decoupling reflects
the fact that the divergence of a traceless symmetric tensorial density of 
rank two is Weyl-invariant in two dimensions.}. Indeed, the diffeomorphism 
constraints,  
\begin{equation}\label{piz}
H_i=-2\sqrt{g}\nabla_{j}\frac{\pi^j_{~i}}{\sqrt{g}}-\sum_n\delta^2 (
x-q_n)P_{ni} =0, 
\end{equation} 
become
$$
\partial_z \pi^{Tz}_{~~~\bar z}=-\frac{1}{2}\sum_n\delta^2 (
x-q_n)P_{n z}
$$ 
{\rm and}
$$
\partial_{\bar z} \pi^{T\bar z}_{~~~z}=-\frac{1}{2}\sum_n\delta^2 (
x-q_n)P_{n\bar z},
$$
which are  solved by
\begin{equation}\label{solpiz} 
\pi^{T\bar z}_{~~~ z}=-\frac{1}{2\pi}\sum_n
\frac{P_{nz}}{z - z_n}+P^{\bar z}_{~z}(z)
\end{equation} 
and its complex conjugate expression. Here we introduced the  complex 
coordinates  $z$ and $\bar z$ with $z=x+iy$ and $P^{\bar z}_{~z} (z)$
is an {\it entire} function. 
In the following,
we shall adopt the {\it maximally slicing} gauge $K=0$, where
$\pi^{ij}$ coincides with its traceless part so that so that eq.(\ref{solpiz})
holds for $\pi^{\bar z}_{\ z}$ and its complex conjugate for
$\pi^{z}_{\ \bar z}$.

We come now to the Hamiltonian constraint,
\begin{equation}
\label{Nvariation} 
H=\frac{1}{\sqrt g}\left[\pi^a_{\ b}\pi^b_{\ a}-(\pi^c_{\
c})^2\right]-\sqrt g R 
\end{equation} 
$$ 
+\sum_n\delta^2(z-
z_n)\sqrt{m_n^2 +P_{ni}~g^{ij}P_{nj}}=0.  
$$ 
If we integrate the l.h.s of eq.(\ref{Nvariation}) over a slice 
of constant time, an elementary application of the Gauss-Bonnet 
theorem shows that the  condition $K=0$  is consistent only with 
open  universes or with closed universes of genus $0$. In the 
latter case consistency can be achieved only for static universes
as it will be shown in Sect.4. Moreover, for open universes,
eq.(\ref{Nvariation}) determines the asymptotic behavior of the  
momenta $\pi^{i}_{~j}$  in terms of that of $\sigma$. In fact 
if we assume that the asymptotic metric has to be conical, 
$e^{2\sigma}\sim (z\bar z)^{-\mu}$,~ $\mu>0$, the above equation implies
that 
$$
\int d^2z ~2\pi^z_{~\bar z} \pi^{\bar z}_{~z} \exp(-2\sigma)<\infty.
$$
This, in turn, entails that $\pi^{\bar z}_{~z}$ at infinity vanishes
faster than $1/z$. Thus, the requirement 
of having a conical metric at infinity sets the function 
$P^{\bar z}_{~z}(z)$ to zero  and in addition it imposes that 
$\sum_n P_{nz}=0$.
Because of the explicit form of the solution (\ref{solpiz}), 
we actually end up with a $\pi^{\bar z}_{~z}$ that goes to zero 
as $1/z^2$. As a consequence, if we define, mimicking the four 
dimensional counterpart, the total momentum 
$P_{tot}$ of the system as the surface integral of
$\pi^{i}_{~j}$,  we are led to conclude that $P_{tot}$
vanishes identically. It is worth noticing that in 
the present  gauge $P_{tot}$ is simply equal to 
$\sum_n {P_{n}}$  and no gravitational correction appears.
The  clash in $(2+1)-$dimensions between the boundary conditions ensuring
finite energy and those ensuring finite momentum was first noticed
in \cite{Henn84,Des85,ashtekar}. In particular, in \cite{Des85} the origin
of this conflict was traced back to the impossibility of performing
asymptotic boosts without introducing spurious singularities in the metric.

We return now to eq.(\ref{Nvariation}). It is not difficult to show
\cite{pmdsann} that $e^{-2\sigma}$ vanishes on the particle
singularities and as a  
result the hamiltonian constraint assumes the form, after setting
$2\tilde\sigma = 2\sigma -\ln(2\pi^{\bar z}_{\ z} \pi^z_{\ \bar z})$
\begin{equation}
\label{eqsigmatilde}
2\Delta\tilde\sigma=
\end{equation}
$$
-e^{-2\tilde\sigma}-\sum_B
4\pi\delta^2(z- z_B)+\sum_n\delta^2(z- z_n) m_n. 
$$
The above equation is an inhomogeneous Liouville equation. The sources
are given by the particles and by some other auxiliary sources whose
position $z_B$ are the zeros of $\pi^{\bar z}_{\ z}$ and their strength 
is fixed to $-4\pi$. The  conformal factor $\sigma$ is actually regular
at the points $z_B$ and for this reason we shall refer to them as 
{\it apparent} singularities.

We remark that
the reduction of the hamiltonian constraint to a Liouville equation is
the result of the $K=0$ condition. The $K={\rm const}\neq 0$ gauge
gives rise, as seen from eq.(\ref{Nvariation}) to an equation of the
sinh-Gordon family which is of much more complex nature.

Finally we note that the conformal gauge i.e. $g_{ij}=
e^{2\sigma}\delta_{ij}$ is still not a complete gauge fixing. One is
allowed to perform analytic transformation on $z$ which leave, in the
case of open universe, the point at infinity fixed i.e. time dependent
translations, rotations and dilatations. This freedom can be exploited
e.g. to keep one particle fixed at $z_1=0$ and an other fixed at
$z_2=1$. In general such a fixing is equivalent to a frame which
rotates at infinity. 

\section{Equations for the lapse and shift functions}

Before discussing the solutions of the inhomogeneous Liouville
equation we shall write down the equations for $N$ and $N^i$; these
are an outcome of requiring that the gauge conditions are preserved 
in the Hamiltonian evolution. Combining the canonical equation for 
$\dot\pi^{ij}$ and the one for $\dot g_{ij}$ the relation $K=0$ 
(or equivalently $\pi^i_{\ i}=0$) takes the form
\begin{equation}
\label{eqN2}
\Delta N = e^{-2\tilde\sigma} N.
\end{equation}
Again we have exploited the vanishing of $e^{-2\sigma}$ at the
particle sources \cite{pmdsann}. Since we have chosen the
conformal gauge, the  
traceless part of the equation for $\dot g_{ij}$ determines $N^i$,
{\it  i.e.}
\begin{equation}\label{Nzequation}
\partial_{\bar z}N^{z}=- \pi^{z}_{\ \bar z} ~e^{-2\sigma} N
\end{equation}
and its complex conjugate. These complete the list of our equations
for the metric. By applying Gauss theorem to eq.(\ref{eqN2}) one 
easily derives the
asymptotic behavior of $N$ for large $|z|$
\begin{equation}\label{behavior}
N\sim \ln(z\bar z).
\end{equation}
Such an asymptotic behavior of the lapse function is typical of the
instantaneous York gauge and should be contrasted with the behavior
$N\rightarrow {\rm const.}$ of the familiar DJH gauge \cite{DJH}. The
behavior (\ref{behavior}) holds both when the plane describes the time
slice of
an open universe and when it describes the sphere; but in the latter 
case in order for the scalar $N$ to be well defined on the Riemann
sphere is should go to a constant at infinity. This occurs only for
$\pi^i_{\ j}\equiv 0$ i.e. when the equation for $N$ reduces to $\Delta
N=0$. Thus we see that except for the trivial stationary case in which
$P_n=0$ the York instantaneous gauge applies only to open universes.

\section{The Liouville equation}
As is well known, the problem of solving the inhomogeneous Liouville 
equation (\ref{eqsigmatilde})  can be reduced to that of solving
a fuchsian second order differential equation of
projectively canonical type \cite{kra}
\begin{equation}\label{Qequ} 
y''(\zeta) + Q(\zeta) y(\zeta)=0
\end{equation}
where 
\begin{equation}
\label{Q(z)}
Q(z) = \sum_n \left[\frac{1-\mu_n^2}{4(z-z_n)^2} +
\frac{\beta_n}{2(z-z_n)}\right] 
\end{equation} 
$$
+ \sum_B \left[ -\frac{3}{4(z-z_B)^2}
+\frac{\beta_B}{2(z-z_B)}\right].  
$$
If we define the function $f(z)$ as the ratio of two independent
solutions of eq. (\ref{Qequ}),
\begin{equation} 
\label{fdiffeo}
f(z) = \frac{k y_1(z)}{y_2(z)},
\end{equation}
the solution of eq. (\ref{eqsigmatilde}) is  built from the Poincar\'e
metric for the pseudosphere by performing on it the conformal diffeomorphism
defined by eq. (\ref{fdiffeo})
\begin{equation}\label{fexpression}
e^{-2\tilde\sigma} =\frac{8 f'(z) \bar f'(\bar z)}{(1-f(z)\bar f(\bar z))^2}.
\end{equation}
The poles in (\ref{Q(z)})  include both particle and apparent
singularities. The residues of the second order poles in $Q(z)$
are directly related to the strength of the delta singularities 
in eq.(\ref{eqsigmatilde}): in fact, $\mu_n = m_n/4\pi$ while the
second order residue at  the apparent singularities  is fixed 
to be $-3/4$. The latter value is required by the regularity of
$2\sigma$ at $z_B$. The determination 
of the first order residues is more difficult and it is related 
to the knowledge of the explicit solution of the Riemann-Hilbert 
problem associated to (\ref{Qequ}). They are fixed in principle 
by the requirement that the conformal factor $\sigma$ be monodromic 
in the complex plane. As eq.(\ref{fexpression}) can be rewritten 
in the form 
\begin{equation} \label{wronsk}
e^{-2\tilde\sigma} = \frac{8 w_{12}\bar w_{12}}{(y_2\bar y_2 - 
k\bar k ~ y_1 \bar y_1)^2} 
\end{equation}
where $w_{12}$ is the constant wronskian,  it is seen that such a 
requirement is equivalent to the imposition that the vector $(k y_1,y_2)$ 
transforms under a representation of $SU(1,1)$ when one encircles 
a singularity in the complex plane. It provides the relation between 
the geometric structure approach \cite{Thurston} and the one here 
described. 

An important feature in the solution of eq.(\ref{eqsigmatilde}) 
is that one has to give the residue of the singularity at 
infinity and this information is not contained in the parameters 
appearing in the sources. In different words, in solving
eq.(\ref{eqsigmatilde}) the total energy of the system has to be  
specified as an arbitrary parameter $M$ because  the positions 
and the momenta of the particles at the initial time are not enough to 
determine its value.  The above conclusion might seem counterintuitive  
especially if we compare it  with what happens in the DJH {\it geometrical}
approach, where the knowledge of all $z_i$ and $p_i$ fixes $M$ completely
through a nonlinear composition rule for the momenta.
This apparent mismatch simply reflects their choice to have,
in contrast with the present approach \cite{BCV,welling,pmdsann},  
a {\it flat}  (not globally defined) 
metric. As a consequence, in their approach, all the dynamical 
information is  shifted into the  particle variables.
In our approach a complete set of initial conditions for the conformal 
factor $\tilde \sigma$ is instead provided by any unconstrained 
parameterization{\footnote{
In fact the $\beta_n$ and $\beta_A$ are
not all independent: In the next section we shall see that fuchsianity
and the value of the residue at infinity impose two sum rule on them.
Moreover, the requirement that conformal factor $\sigma$ be regular 
at  $z_A$ fixes the $\beta_A$ in terms of the other parameters by 
means of the so-called {\it no-log} condition, see sec.9.}
of the rational function $Q(z)$: an example will
be given in sec. 9.

The conformal factor $2\tilde\sigma$ plays the key role in all
subsequent developments; once the solution of eq.(\ref{eqsigmatilde})
is known the 
equations for $N$ and $N^i$ are solved as follows. As the sources in
eq.(\ref{eqsigmatilde}) do not depend on the total energy $M$, by taking the
derivative with respect to $M$ we find a solution to the homogeneous
equation (\ref{eqN2}) for $N$ i.e. apart for an arbitrary normalization
constant we have
\begin{equation}
\label{Neq}
N=\frac{\partial (-2\tilde\sigma)}{\partial M}.
\end{equation} 
As for the equation (\ref{Nzequation}) multiplying it by $2
\pi^{\bar z}_{\ z}$ and 
using eq.(\ref{eqN2}) and taking into account that $\pi^{\bar z}_{\ z}$ is a 
function of $z$ we get
\begin{equation}\label{Nzexp}
N^z =-\frac{2}{\pi^{\bar z}_{\ z}(z)} \partial_z N +g(z)
\end{equation}
where $g(z)$ is a function of $z$. A similar solution holds
for $N^{\bar z}$. The function $g(z)$ has to be chosen as to cancel the poles 
generated by the zeros of $\pi^{\bar z}_{\ z}(z)$, which occur only in
presence of three or more particles, and if we are interested in
describing a reference system which 
does not rotate at infinity $g(z)$ has to be chosen as to give $N^z <
|z|$ at infinity. We shall see in sect.7 that,  due to
$\sum_n P_{nz}=0$,  such a function can diverge at infinity at most
linearly. As we 
shall see $g(z)$ encodes the time evolution of all the quantities of
the problem.

\section{The particle equations of motion}
The canonical equations for the particle positions $z_n$ and momenta
$P_n$ are obtained by varying the action with respect to $z_n$ and
$P_n$ and are given by
\begin{equation}\label{dotz}
\dot z_n = -N^i(z_n,t) = - g(z_n),
\end{equation}
\begin{equation}\label{dotP}
\dot P_{n z} = P_{n a}\frac{\partial N^a}{\partial z} -m_n
\frac{\partial N}{\partial z}
\end{equation}
and their complex conjugate expressions. We note that $\dot z_n$ is
simply given in terms of the meromorphic function $g(z)$ as in
eq.(\ref{Nzexp}) $\pi^{\bar z}_{\ z}$ diverges at the particle
positions while $\partial_z N$ stays finite. 
It is of interest that the r.h.s of eq.(\ref{dotP}) depends
only on the function  $g(z)$ and on the linear residue $\beta_n$ 
at the particle singularity. In fact by using the behavior of the solutions
eq.(\ref{Qequ}) at the particle singularities, one obtains 
\begin{equation}\label{dotP2}
\dot P_{n z} = 4\pi \frac{\partial \beta_n}{\partial
M}+P_{nz} g'(z_n).
\end{equation}
The fact that the time evolution is governed by the function $g(z)$ will
be the common denominator of the following developments.

\section{Conservation laws}
For a fuchsian differential equation of type (\ref{Qequ}) the following
Fuchs relations hold \cite{yoshida}
$$
\sum_n\beta_n +\sum_B\beta_B=0; ~~~~~
1-\mu ^2_\infty = 
$$
\begin{equation}\label{fuchsrelations}
=\sum_n(1-\mu _n^2
+2\beta_n z_n) + \sum_B (-3 +2\beta_B z_B).
\end{equation}
The former expresses the fact that the function $Q(z)$ behaves as $1/z^2$
at infinity ({\it Fuchs condition}); the latter provides the second
order residue of $Q(z)$ at infinity ({\it the total mass} of the system) 
in  terms of the other parameters.

Below, we shall shows that these conditions implie some non-trivial 
conservation laws. We start 
recalling  the equation of motion for $\pi^{\bar z}_{~z}$
\begin{equation}
\label{dotpi}
\dot \pi^{\bar z}_{\ z}=
2 e^{2\sigma}\partial_z(e^{-2\sigma}\partial_z
N)+2\pi^{\bar z}_{\ z}\partial_z N^z + N^z\partial_z\pi^{\bar z}_{\ z} 
\end{equation}
which by means of the eqs.(\ref{Neq}) and (\ref{Nzexp}) can be recasted 
as follows
\begin{equation}
\label{eqpi1}
\dot \pi^{\bar z}_{~ z}=-4{\partial_M Q(z)}+
\partial_z \pi^{\bar z}_{~z} g+2 \pi^{\bar z}_{~z}\partial_z g.
\end{equation}
We have used a standard result about the Liouville equation that permits
to connect $\tilde\sigma$ to $Q(z)$
\begin{equation}\label{emt}
Q(z) =
-\frac{1}{2}[\partial^2_z(2\tilde\sigma)+\frac{1}{2}(\partial_z
(2\tilde\sigma))^2].
\end{equation}
In a field theoretical language $Q(z)$ is proportional to the energy momentum
tensor associated to the Liouville equation.

If we integrate both sides of eq.(\ref{eqpi1}) over a closed path in 
the complex plane that encircles all the singularities, we find
\begin{equation}\label{const1}
\frac{d}{d t} \sum_n P_{nz} = 4\pi \partial_M (\sum_n \beta_n+\sum_B
\beta_B)-
\end{equation}
$$
2\pi\oint dz\left ( \partial_z 
\pi^{\bar z}_{~z} g+2 \pi^{\bar z}_{~z}\partial_z g\right ).
$$
Since both the sum over momenta and that  over the first order residues
vanish, the  remaining integral must vanish as well. Recalling that 
$\pi^{\bar z}_{~z}\simeq 1/z^2$ for large radius, we must conclude 
that $g$ can grow at most linearly at infinity. Moreover, if we 
combine this  information with the fact that the only singularities of 
$g$ at finite radius can be poles, we find that $g$ must be a rational
function.

Next, we multiply eq.(\ref{eqpi1}) by $z$ and we integrate again 
over a closed path in the complex plane to obtain
\begin{equation}
\label{cons2}
\frac{d}{dt}\sum_n z_n P_{nz}=
\end{equation} 
$$
4\pi\partial_M 
(\sum_n z_n \beta_n+\sum_B z_B \beta_B) = 1-\frac{M}{4\pi}. 
$$
The previous equation combined with its complex conjugate gives both the law
of conservation of angular momentum  
\begin{equation}
\dot L = \frac{d}{dt}  \sum_n i (z_n P_{nz} - \bar z_n P_{n \bar z})
=0
\end{equation} 
and the ``generalized conservation law'' 
\begin{equation}\label{dilatation}
\frac{d}{dt}\sum_n (z_n P_{nz} + \bar z_n P_{n \bar z})=2(1-\frac{M}{4\pi}).
\end{equation}
The fact that the terms depending on $g$ in eq.(\ref{eqpi1}) do not 
contribute can be easily seen if we write the r.h.s of (\ref{eqpi1}) 
times $z$ in the form 
\begin{equation}
-4 z\partial_M Q(z) + \pi^{\bar z}_{~z} (z\partial_z g-g)+
\partial_z (z \pi^{\bar z}_{~z} g).
\end{equation}
The last term is harmless since it is a total derivative of a rational
function; the previous one 
falls again as $1/z^2$ since the combination $(z\partial_z g-g)$ cancels
the linear behavior of $g$ at infinity.

Eq.(\ref{cons2}) is a severe constraint on the form of the reduced
hamiltonian ${\cal H}$ describing the motion of the particles when 
gravity is eliminated. In fact $\sum_nP_{nz}=0$
imposes 
\begin{equation}
{\cal H} = {\cal H}(\zeta_2,\dots \zeta_{\cal N}, P_2, \dots P_{\cal N})
\end{equation}
with $\zeta_n = z_n - z_1$ and eq.(\ref{cons2}) imposes that 
\begin{equation}
\sum_{n>1} \left (P_{nz}\frac{\partial{\cal H}}{\partial P_{n z}}-  
\zeta_{n}\frac{\partial{\cal H}}{\partial \zeta_{n}}\right ) =
1-\frac{M}{4\pi}.  
\end{equation}
The general solution of this equation has the form  
\begin{equation}
{\cal H}= \sum_{n>1}\left ( \log(P_{nz} P_{n\bar 
z})+ \log(\zeta_{n} \bar \zeta_{n})^{\alpha_n}\right)+ {\cal H}_0
\end{equation}
where ${\cal H}_0$ is a homogeneous function of degree zero in the variables
$P_{nz}$ and $1/\zeta_n$ and $\sum_{n>1} (1-\alpha_n)= 1-\frac{M}{4\pi}$. 
This property will be seen explicitly realized for two 
particles with ${\cal H}_0=0$.

\section{The two body problem}
In the two body problem, $\pi^{\bar z}_{\ z}$ has no zero i.e. we are in
absence of apparent singularities; $g(z) = z/P_{1z}(z_2-z_1)$ in order
to have a non rotating frame at infinity. The simple residues $\beta_1,
\beta_2$ are provided by the Fuchs relations in terms of particle 
masses $m_1, m_2$ and the total energy. $f$ appearing in
eq.(\ref{fexpression}) is given by the ratio of two hypergeometric
functions which provides the metric \cite{BCV,welling,pmdsann}. On the
other hand it is 
interesting that the particle trajectories can be determined even
before the explicit computation of the metric. In fact we have
for the two particles the equation of motion (\ref{dotz},\ref{dotP}). The rewriting
of the two formulas for the problem at hand gives for $l=z_1-z_2$ and
$P_z = P_{1z}=-P_{2z}$
\begin{equation}\label{hameq}
\dot l =\frac{1}{P_z},~~~~\dot P_z = - \frac\mu {l}
\end{equation}
whose general solution is
\begin{equation}\label{trajectory}
l = {\rm const}~ [(1-\mu )(t-t_0) - iL/2]^{\frac{1}{1-\mu }}.
\end{equation}
The scattering angle is immediately computed from the previous
equation to be $\theta_{scatt} = \pi M/(4\pi -M)$ \cite{DJH,hooft3}. It is
interesting to note that the scattering angle 
depends only on the total energy and not on the masses of the
individual particles.
The two equations (\ref{hameq}) are generated by the hamiltonian
$H_{eff}= \ln (P^2_x + P^2_y) + \mu \ln (l^2_x + l^2_y)$ which is
unique up to an additive constant. Such an hamiltonian should not be
confused with the total energy of the system; it is the effective
hamiltonian for the time development of the two particle system and
contains as an essential parameter the total energy $M = 4\pi \mu$. 

\section{The ${\cal N}$ body problem}
We shall consider in this section the ${\cal N}$-particle case with
particular attention to ${\cal N}=3$. 
In flat space the number of Lorentz invariants is easily seen to be
$3 {\cal N}-3$
as for $n>3$ the $2+1$-vector ${\bf P}_n$ is determined by the scalar
product of ${\bf P}_n$ with ${\bf P}_1, {\bf P}_2, {\bf P}_3$. In
curved space we have the same number of invariants. As it is well
known here the invariants are replaced by the traces of the holonomies
around the world lines of an arbitrary collection of particles
\cite{DJH,carrol,menottiseminara}. The
parallel transport monodromy matrices which belong to $SO(2,1)$ or in
the fundamental representation to $SU(1,1)$ are defined up to a
conjugation.  
Keeping in mind that every $SU(1,1)$ holonomy has three real degrees
of freedom we have $3{\cal N}$ degrees of freedom to which we have to
subtract the three degrees of freedom of the $SU(1,1)$ conjugation
thus reaching as expected the same number of invariants as in the flat
case. 

In the present approach parameterizing the dynamics in terms of the
position of the particles and their momenta is not a convenient 
choice, because these, as we have anticipated, are not the natural 
variables for the associated Riemann-Hilbert problem.
In order to understand the problem better let us consider the case
${\cal N}=3$. It is useful to go over from the equation in projectively
canonical form (\ref{Qequ}) to the equivalent one presented below
\cite{yoshida}
\begin{equation}\label{equivalent} 
y''+p y' + q y =0
\end{equation} 
with
\begin{equation}\label{Qequivalent}
Q = q- \frac{p'}{2} -\frac{p^2}{4}. 
\end{equation} 
The transformation can be so chosen as to have the following
Riemann scheme
\begin{equation}
\left (
\begin{array}{ccccc}
0&1&z_3&z_A&\infty\\
0&0&0&0&\rho_\infty\\
\mu _1&\mu _2&\mu _3& 2&\rho_\infty+\mu _\infty
\end{array}\right ) 
\end{equation}
where the Fuchs relation 
\begin{equation} 
\mu _1+\mu _2+\mu _3+\mu _\infty +2\rho_\infty=1
\end{equation} 
fixes $\rho_\infty$ in terms of the other parameters and 
$$
p(z) =
\frac{1-\mu _1}{z}+\frac{1-\mu _2}{z-1}+\frac{1-\mu _3}{z-z_3}-\frac{1}
{z-z_A}; 
$$
$$ 
q(z) =
\frac{\kappa}{z(z-1)}-\frac{z_3(z_3-1) H }{z(z-1)(z-z_3)}+
$$
$$
\frac{z_A(z_A-1) b_A}{z(z-1)(z-z_A)} 
$$
with $\kappa = \rho_\infty(\rho_\infty+\mu_\infty)$.
Here we use the unconstrained parameterization of ref.\cite{yoshida};
the connection with the parameters $\mu_n$ and $\beta_n$ can be easily
obtained by comparing the residues in eq.(\ref{Qequivalent}).

The parameters
$\mu _i$ and $\mu _\infty$ are free real parameters which give the
trace of the holonomies around the particles and infinity (total
energy). $z_A$ is given in terms of the $z_n$ and the $P_n$
($n=1,2,3$), while $H$ is given by the
no-logarithm condition \footnote{Here $Q_A(z)$ denotes the  
function $Q$ without its poles in $z_A$.}  
$ Q_A(z_A)={\beta_A^2/2}$,
i.e. the requirement that the point $z_A$ is a regular point for
$2\sigma$ \cite{yoshida}. Such $H$ is easily found
\begin{eqnarray}
&&H=\frac{z_A(z_A-1)(z_A - z_3)}{z_3(z_3-1)}\left\{b_A^2 - (\frac{\mu _1}{z_A}
+\right.\nonumber\\
&&\label{garnierhamiltonian}\left.
+\frac{\mu _2} {z_A-1} +\frac{\mu _3-1}{z_A-z_3})b_A +
\frac{\kappa}{z_A(z_A-1)} \right\}. 
\end{eqnarray} 
Thus we have the following free real parameters
$\mu _1, \mu _2, \mu _3, \mu _\infty, {\rm Re}(z_A),{\rm Im}(z_A), {\rm
Re}(b_A),{\rm Im}(b_A)$ which are eight real parameters, to be
contrasted with the six real 
invariants we have enumerated at the beginning of this section. The
reason is that the constraint that the holonomies described by
eq.(\ref{equivalent}) belong to $SU(1,1)$, fixes the value of ${\rm
Re}(b_A),{\rm 
Im}(b_A)$ in terms of all other parameters. The explicit form of this
relation has been investigated in the mathematical literature
\cite{schlesinger} but we
are not aware of any explicit form of it. Adding a new particle
introduces a new particle singularity with exponent $\mu _4$
i.e. $p(z)$ acquires a new term of the type $(1-\mu _4)/(z-z_4)$,
but in addition a new apparent singularity is generated say at
$z_B$. The indices of this new singularity have to be integers,
$0$ and $2$ while the new simple residue $b_B$ has to be a function of
all other parameters, and in particular of $z_B$, because the
$SU(1,1)$ character of the holonomies allow only an increase of $3$ in
the number of free real parameters i.e. $\mu _4, {\rm Re}z_B, {\rm
Im}z_B$. Adding other particles does not alter the procedure.

The complete solution of the dynamical problem requires that we provide
the equations of motion for all the variables parameterizing the dynamics
of the system and in particular for the apparent singularities and their 
first order residues. To this purpose,  we shall follow  here a
synthetic approach which will encode in a single equation all the
information on the time development of the positions and residues of
all singularities. To accomplish this we shall exploit the trace part of
equation for $\dot g_{ij}$  i.e.
\begin{equation}\label{dotsigma3}
\frac{d(2\sigma)}{d t} = N^z\partial_z(2\sigma) + \partial_z N^z +
                          N^{\bar z}\partial_{\bar z}(2\sigma) +
                          \partial_{\bar z} N^{\bar z} 
\end{equation} 
which combined with the time development of $\pi^{\bar z}_{\ z}(z)$,
eq.(\ref{dotpi}),
provides the time derivative of the reduced conformal factor
$\tilde\sigma$ i.e.
\begin{equation}
2 \dot{\tilde\sigma} = g(z)\partial_z (2\tilde\sigma)- g'(z) + {\rm c.c.}
\end{equation}
We shall exploit now  the fundamental relation (\ref{emt}) of the theory of 
fuchsian differential equation \cite{yoshida,kra}. 
Taking the time derivative of eq.(\ref{emt}) we have
\begin{equation}
\label{Qequation}
\dot Q(z) = \frac{1}{2}g'''(z) + 2 g'(z) Q(z) +g(z) Q'(z).
\end{equation}
The above equation implies that the time development of $Q(z)$ is
given by a conformal 
diffeomorphism generated by $Q(z)$ itself. We are not in presence
of a classical 
transformation but of an anomalous one. Extensions of the classical
Liouville theory have been considered in \cite{dhokerjackiw}.

It is interesting that this equation, in addition to the
particle equations of motion, contains the whole Garnier system for
the time evolution of the apparent 
singularities.  We recall that $g(z)$ is the rational function introduced in
eq.(\ref{dotP}); it is given by a sum of simple poles located at the
apparent singularities and of a first order polynomial.

Equating the simple residue of eq.(\ref{Qequation}) at infinity  one simply
obtains 
\begin{equation}
\frac{d}{dt}(\sum_n\beta_n +\sum_B\beta_B)= 0
\end{equation}
which is already contained in the Fuchs relation (\ref{fuchsrelations}), while
equating the second order residues at infinity we obtain
$$
\frac{d}{dt}\left[\sum_n(1-\mu_n^2) +2 \sum_n \beta_n z_n +
2\sum_B\beta_B z_B\right]= 0 
$$
which using the second Fuchs relation (\ref{fuchsrelations}) provides
$d\mu^2_\infty/dt=0$, where $\mu_\infty = 1- M/4\pi$ i.e. the conservation
of total energy. On the other hand if we equate the second order
residues of eq.(\ref{Qequation}) at the particle pole $z_n$ we obtain
$d\mu^2_n/dt=0$, where $\mu_n = m_n/4\pi$ i.e. the conservation of the
individual particle rest masses.
Matching of the third order residue on the particle singularity $z_n$
gives
\begin{equation}
\dot z_n = -g(z_n)
\end{equation}
while matching of the first order residue gives $\dot \beta_n$ i.e.
\begin{equation}
\dot \beta_n = \frac{1-\mu^2_n}{2}g''(z_n)+g'(z_n)\beta_n.
\end{equation}
The motion of the apparent singularities is given by the matching of
the residues of the third order pole at $z_A$. If we Laurent expand
$g(z)$ around $z_A$ 
\begin{equation}
g(z) = \frac{g_{-1}}{z-z_A} + g_0 +g_1 (z-z_A) +
O((z-z_A)^2) 
\end{equation}
we obtain
\begin{equation}
\label{zadot}
\dot z_A = -g_{-1}\beta_A - g_0
\end{equation}
while the first order pole gives
\begin{equation}
\label{betaadot}
\dot\beta_A = -2 g_{-1} Q'_A(z_A) + g_1 \beta_A - \frac{3}{2}g_2.
\end{equation}
Equations (\ref{zadot}) and (\ref{betaadot}) can be written in
hamiltonian form; for 
example in the three body problem, by choosing $g(z)$ as to cancel the
pole in eq.(\ref{Nzexp}) and keep particle $1$ fixed at
$z_1=0$ and particle $2$ fixed at $z_2=1$, we have
$$
\frac{\partial z_A}{\partial z_3 } = \frac{z_A (z_A -1)(z_A
-z_3)}{z_3(z_3-1)}\times  
$$
\begin{equation}
\left[ \beta_A  
-\frac{1}{z_A-1}-\frac{1}{z_A}\right]= \frac{\partial H}{\partial b_A}
\end{equation}
and
\begin{equation}
\frac{\partial \beta_A}{\partial z_3}=  -\frac{\partial H}{\partial z_A}.
\end{equation}
where $H$ is the so called Garnier hamiltonian and it given by
(\ref{garnierhamiltonian}). The parameter $\beta_A$ is connected to
those appearing in $H$ by the relation  
with
\begin{equation}
\frac{\beta_A}{2} = b_A +\frac{1}{2}\left( \frac{1-\mu _1}{z_A}
+\frac{1-\mu _2}{z_A-1}+ \frac{1-\mu _3}{z_A-z_3}\right). 
\end{equation}
The explicit expression of $H$ can also be written for the general
$\cal N$ body problem. 

It is now possible to interpret the problem of gravity in presence of
particles \`a la Einstein- Infeld- Hoffmann \cite{ehi} i.e. as pure gravity
without particles in which one allows for singular solutions of the
field equations. What we found is that the deficit of the singularity
has to remain constant (equivalent to the constancy of rest masses),
the total angular deficit has to remain constant (equivalent to the
constancy of the total energy) and the field singularities follow
geodesics of space- time. 

\section{Conclusions}

The application of the instantaneous York gauge $K=0$, to $2+1$
dimensional gravity is practically restricted to the case of open
universes; on the other hand in this case the advantages are several
and profound. The momenta $\pi^a_{~b}$ conjugate to the space metric
have only two independent components and the diffeomorphism constraint
has a 
very simple solution i.e. the $\pi^{\bar z}_{~z}$ ($\pi^{z}_{~\bar
z}$) are rational functions of $z$ ($\bar z$) whose residues
are the particle momenta $P_n$.

The advantage is still greater with regard to the hamiltonian
constraint which in this gauge is equivalent to an inhomogeneous
Liouville equation for the reduced conformal factor $\tilde\sigma$; 
the sources are the particle singularities i.e. the poles of
$\pi^{\bar z}_{~z}$ and the apparent singularities, i.e. the zeros of
$\pi^{\bar z}_{~z}$. The function $N$ is given by the derivative of 
$\tilde\sigma$ with respect to the total energy while $N^z$ is given
in terms of $N$ by
\begin{equation}
N^z = -\frac{2}{\pi^{\bar z}_{~z}}\partial_z N +g(z),
\end{equation}
where $g(z)$ is a rational function which cancels the polar
singularities of the first term and grows at infinity not faster than
$z$. The linear term in $g(z)$ is fixed if we want to deal with a
reference frame which does not rotate at infinity. 

The equations of motion for the position of the particles and of the
apparent singularities and also the time dependence of the linear
residues at such singularities are given by the time development
of the function $Q(z)$ entering the fuchsian differential equation
which underlies the conformal factor describing the space metric. 
Such a time development is given by the energy momentum
tensor of a conformal Liouville theory and it encodes completely the
dynamics of the system.

\section{Acknowledgments}

We are grateful to Marcello Ciafaloni, Stanley Deser, Marc Henneaux
and Roman Jackiw for useful discussions.

\end{document}